\documentclass[aps,pre,reprint,showpacs,superscriptaddress,groupaddress,showkeys,floatfix,byrevtex]{revtex4-1}
\usepackage{amsmath}
\usepackage{amsfonts}
\usepackage{amssymb}
\usepackage{graphicx,xcolor,subeqnarray}
\usepackage{hyperref}
\usepackage{mathtools}
\usepackage{soul}
\begin{document}
\title{Individual particle persistence antagonizes global ordering in populations of nematically-aligning self-propelled particles}

\author{J. M. Nava-Sede\~no}
\email{manikns@ciencias.unam.mx}
\affiliation{Departmento de Matemáticas, Facultad de Ciencias, Universidad Nacional Aut\'onoma de M\'exico, Circuito Exterior, Ciudad Universitaria, 04510 Ciudad de M\'exico, M\'exico}

\author{R. Klages}
\affiliation{Centre for Complex Systems, School of Mathematical Sciences, Queen Mary University of London, Mile End Road, London E1 4NS, United Kingdom}

\affiliation{London Mathematical Laboratory, 8 Margravine Gardens, London W6 8RH, United Kingdom}

\author{H. Hatzikirou}
\affiliation{Technische Universit\"at Dresden, Center for Information Services and High Performance Computing, N\"othnitzer Stra{\ss}e 46, 01062 Dresden, Germany}
\affiliation{Mathematics Department, Khalifa University, PO Box 127788, Abu Dhabi, United Arab Emirates}

\author{Francisco J. Sevilla}
\email{fjsevilla@fisica.unam.mx}
\affiliation{Instituto de F\'isica, Universidad Nacional Aut\'onoma de M\'exico, Apdo.\ Postal 20-364, 01000, Ciudad de M\'exico, M\'exico}

\author{A. Deutsch}
\affiliation{Technische Universit\"at Dresden, Center for Information Services and High Performance Computing, N\"othnitzer Stra{\ss}e 46, 01062 Dresden, Germany}

\date{\today}

\begin{abstract}
The transition from individual to collective motion plays a significant role in many biological processes. While the implications of different types of particle-particle interactions for the emergence of particular modes of collective motion have been well studied, it is unclear how particular types of individual migration patterns influence collective motion. Here, motivated by swarming bacteria \emph{Myxococcus xanthus}, we investigate the combined effects of the individual pattern of migration and of particle-particle interactions, on the emergence of collective migration.
We analyze the effects of a feature of individual pattern migration, the persistence of motion, on the collective properties of the system that emerge from interactions among individuals; in particular, when nematic velocity alignment interaction mediates collective dynamics. 
We find, through computer simulations and mathematical analysis, that an initially disordered migratory state can become globally ordered by increasing either, the particle-particle alignment interaction strength or the persistence of individual migration. In contrast, we find that persistence prevents the emergence of global nematic order when both persistence and nematic alignment are comparatively high. We conclude that behavior at the population level does not only depend on interactions between individuals but also on the individuals' own intrinsic behavior.
\end{abstract}

\maketitle

\section{Introduction}

The emergence of collective motion, characterized by a phase with
long-range orientational order, originating from local interactions
among individuals, has been the subject of intense research within a
variety of mathematical frameworks in different fields, from biology \cite{vicsek2012collective,MehesIntegrativeBio2014,GilPLOSCompBiol2015,WuQuantBio2015,be2019statistical} to
physics
\cite{VicsekPRL1995,TonerPhysRevLett1995,RaymondPhysRevE2006,AldanaPhysRevLett2007,DossettiPhysRevLett2015},
and lately in systems of robots
\cite{calovi2018disentangling,starruss2012pattern,alfonso2017biology,correll2006collective,shklarsh2011smart}.
In biological systems, pattern formation is intimately related to
  active matter \cite{Rama10}, which consists of a large number of
  self-propelled, called active, particles \cite{Schw03,RBELS12} that
  are interacting with each other.  While diverse patterns of active
motion of biological organisms, and of artificial active particles,
have been observed in isolation \cite{BeDiL16}, mathematical models
for the analysis of collective migration usually assume that in a
dilute regime, individuals perform rather simple persistent
random walks
\cite{plank2012models,szabo2006phase,peruani2008mean}. Meanwhile
  different, more advanced models of single active particle motion are
  available. All of them exhibit the fundamental property that the
  particles are self-driven, in terms of breaking
  fluctuation-dissipation relations, by generating different types of
  persistent motion. These models are typically formulated by
  (overdamped) Langevin dynamics, hence called {\em active Brownian
    particles} \cite{Schw03,RBELS12,BeDiL16}, where either the
  friction coefficient is a nonlinear function of the velocity
  \cite{SEC94}, the velocity vector exhibits rotational diffusion
  \cite{KMDL14}, or particles are driven by colored (typically
  exponentially correlated) Gaussian noise \cite{HJR07}. In addition there 
  is the class of run-and-tumble particles \cite{Berg08} characterised by intermittent dynamics, which
  requires yet another modification of Langevin dynamics by imposing 
  a statistics of tumble times onto the turning angle distribution \cite{KWPES16}.

Recently, the explicit effects of different types of single-particle
migration strategies on the collective behavior of motion have been
studied for particles interacting through volume exclusion
\cite{gavagnin2018modeling}. However, the role of more
  non-trivial active single-particle migration strategies on the
globally-ordered phases of collective motion is much less known
especially when aligning particle-particle interaction is
considered. This problem is of much current interest, and addressed in
this paper. Some authors have already started to consider a similar
question by analyzing how self-propulsion of run-and-tumble particles
affects the separated phase induced by motility (MIPS)
\cite{ray2023increased}. In Ref.~\cite{SperaPhysRevLett2024} the effects of nematic alignment interactions on the collective behavior of active matter are studied. The authors find a generic mechanism where  fluctuations enhance polar order, thus increasing the persistence length of active motion which affects collective behavior. The results exhibit the mutual influence between nematic interaction and persistence in active matter systems.

Self-propelled particles that align among themselves
  and are subject to noise, have been the starting point for the
  analysis of a set of non-equilibrium collective properties that
  emerge in these systems, and became quickly a fast-growing subfield
  of active matter \cite{ChateAnnuRevConMatPhys2020}.
Recent studies on active nematics and self-propelled
  rods have extensively explored the emergence of complex patterns and
  collective behaviors driven by intrinsic particle properties and
  interaction rules. The behavior of nematically aligning particles,
  characterized by their elongated shapes and tendency to align in a
  head-to-tail fashion, has been a focal point of research due to its
  relevance in understanding biological systems
  \cite{peruani2006nonequilibrium,starruss2007bacterial,peruani2008mean,starruss2012pattern}(see
  \cite{Rama10,marchetti2013hydrodynamics,bar2020self}
  for a review). Previous work has demonstrated that nematic
  interactions can lead to the formation of dynamic structures such as
  bands, swirls, and defects, which are the hallmarks of active
  nematic behavior
  \cite{starruss2007bacterial,peruani2006nonequilibrium,sanchez2012spontaneous,doostmohammadi2018active,zhang2018interplay,ChateAnnuRevConMatPhys2020}.
However, the role of persistence generated by
    realistic active particle motion, particularly how it influences
  the formation and stability of these structures, has been less
  studied. Specifically, models dealing with persistent, active
  nematic particles have not focused on the effect of varying
  persistence on density pattern formation
  \cite{igoshin2001pattern,grossmann2016mesoscale,li2017mechanism,mahault2021long}
  or have not considered persistence explicitly
  \cite{borner2002rippling,ngo2014large,putzig2014phase,prathyusha2018dynamically,nava2023vectorial,del2023dichotomous,sharma2024phases}. Rather, most models in active nematics focus on the dynamics of defect creation and annihilation, or the onset of buckling instabilities  \cite{shi2013topological,decamp2015orientational,putzig2016instabilities,henkes2018dynamical,shankar2019hydrodynamics,patelli2019understanding,vliegenthart2020filamentous,arora2022motile}.

Motivated by the experiments of nematically-interacting rod-shaped
particles and of \emph{Myxococcus xanthus} bacteria, with both
persistent and non-persistent phenotypes
\cite{thutupalli2015directional}, 
and interested in defining a minimal model which captures the essential dynamics of such biological systems, we present in this paper a study of
the effects of persistent active motion on the global orientational
order of a population of aligning particles. Such a population is
modeled mathematically by a specifically designed lattice gas
  cellular automaton \cite{Deutsch2005}, which here consists of a
  system of nematically interacting overdamped active Brownian
  particles driven by exponentially correlated Gaussian noise leading
  to an exponentially decaying velocity autocorrelation function as
  observed, e.g., in cell migration experiments
  \cite{Diet22,nava2017cellular}.

While polar or nematic global order can emerge either by increasing
the intensity of the particle-particle interaction or by increasing
the persistence of active motion, here we find that \emph{global
nematic order} is only possible in populations of weakly persistent
particles. High persistence of the active motion hinders the
appearance of global while allowing local nematic order. This effect
agrees with the emergence of different patterns, such as stacks,
sheets, and streams in wild-type and mutant strains of
\emph{M. xanthus} \cite{thutupalli2015directional}.

The structure of the paper is as follows. In Section \ref{sect:Model}, we define the mathematical model. In Section \ref{sect:numerics}, we characterize the behavior of the model by performing computer simulations and calculating order parameters. In Section \ref{sect:analysis}, we perform a mathematical analysis of the model to gain deeper understanding of its dynamics. In Section \ref{sect:conclusions}, we discuss our results and propose potential extensions and improvements to our methodology.

\begin{figure*}
\includegraphics[width=0.75\textwidth]{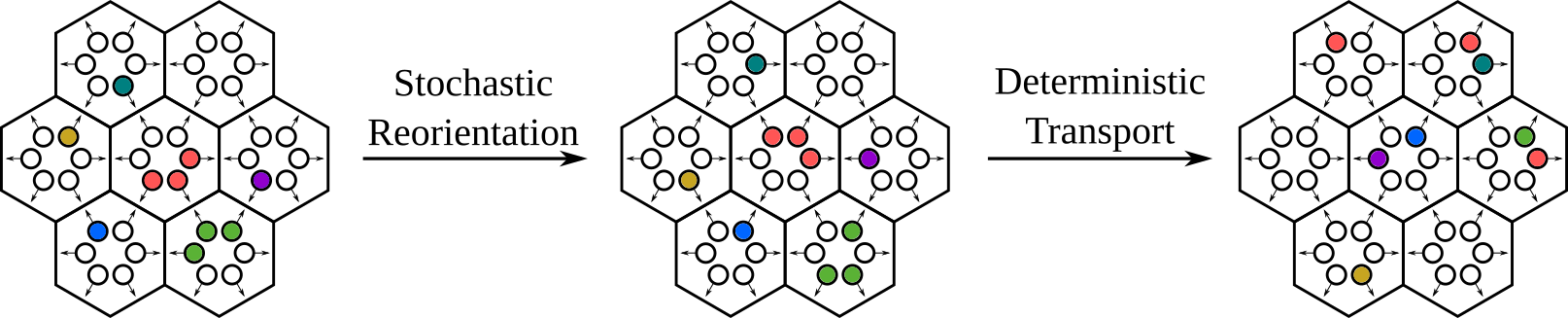}
\caption{Schematic description of the updating process of the system configuration. Each hexagonal lattice site contains six velocity channels (circles) that define the possible direction of motion. Each channel can be occupied by only one agent (occupied channels are depicted as colored circles), the occupation of each channel defining the state of the lattice site. The first step of the updating process involves the selection of a new channel with a transition probability given by Eq.~\ref{trans-real}. In the second step, particles are translocated to the neighboring cell in the direction of the velocity channel chosen in the previous step. Both steps are applied to every single lattice site in parallel at every time step.}
\label{fig:TransitionProcess}
\end{figure*}
\section{\label{sect:Model}Lattice Gas Cellular Automata (LGCA) Model}
In our paper we consider a discrete model in space and time for
studying the dynamics of nematically interacting active Brownian
  particles in the form of a lattice gas cellular automaton
  \cite{Deutsch2005,nava2020modelling,deutsch2021bio}. Here the system micro-states are
specified by the occupation numbers of the six velocity channels
(possible directions of motion) on each site (node) of a
two-dimensional regular triangular lattice $\mathcal{L}$. This
tessellation is preferred since it provides the maximum number of
neighboring nodes on the Euclidean plane, while maintaining
translational invariance, which coincidentally is close to the number
of interacting agents estimated experimentally in some natural systems
exhibiting collective motion \cite{BalleriniPNAS2008}. On each node
$\vec{r}$ of $\mathcal{L}$, the occupancy $s_{\vec{r}}^j$
($j=1,\ldots, 6$) of the velocity channel
$\vec{c}^{j}=(\cos\theta^j,\sin\theta^j)$ with
$\theta^{j}=\frac{\pi}{3}j$, is either empty $s_{\vec{r}}^{j}=0$ (no
particle on node $\vec{r}$ is moving along such direction) or occupied
by at most one particle $s_{\vec{r}}^{j}=1$. In our study, we only consider cases where the number of particles is smaller than the number of sites in the system,thus, the
channel occupancy is Boolean by definition and so $s_{\vec{r}}^{j}\in\{0,1\}$. The
occupation of the $j$-th velocity channel occupation constitutes the
$j$-th entry of the occupation vector
$\vec{s}_{\vec{r}}$. Consequently, the state space of
  the automaton is the set $\mathcal{E}=\{0,1\}^6$.  The particle
density at node $\vec{r}$, $\rho(\vec{r})$, is defined as the fraction
of occupied velocity channels with respect to the total number of
channels, i.e. $\rho(\vec{r})\coloneqq\frac{1}{6}\sum_{j=1}^6
s_{\vec{r}}^j\coloneqq n(\vec{s}_{\vec{r}})/6$, being
$n(\vec{s}_{\vec{r}})$ the number of occupied channels of state
$\vec{s}_{\vec{r}}$.

The dynamics of the system is given by a lattice-gas cellular automaton (LGCA), which consists of synchronous updating rules at each time step. The updating process is carried out in two steps: A stochastic reorientation of the agents direction of motion that updates the occupation of the velocity channels, followed by the translocation of the agent to the neighboring node in the direction of the velocity channel. During the reorientation step, the occupation configuration of each node at time step $\tau$, $\vec{s}_{\vec{r}}(\tau)=\bigl(s_{\vec{r}}^1(\tau),\ldots,s_{\vec{r}}^6(\tau)\bigr)$, is updated to  a virtual state $\vec{s}^{\mathcal{I}}_{\vec{r}}(\tau)=\bigl({s^{\mathcal{I}}}_{\vec{r}}^1(\tau),\ldots,{s^{\mathcal{I}}}_{\vec{r}}^6(\tau)\bigr)$ according to the transition probability $P\bigl(\vec{s}_{\vec{r}}\rightarrow{\vec{s}^\mathcal{I}_{\vec{r}}}\bigr)$, which takes into consideration the particle-conserving ``interaction'' with neighboring nodes as is explained later. Afterwards, the information of the occupied channels $\vec{c}^j$, given by the virtual occupancy state $\vec{s}^\mathcal{I}_{\vec{r}}$, is used to update the system state $\vec{s}_{\vec{r}}(\tau+1)=\bigl(s_{\vec{r}}^1(\tau+1),\ldots,s_{\vec{r}}^6(\tau+1)\bigr)$ by translocating each occupancy $s_{\vec{r}}^j$ at $\vec{r}$, to the corresponding neighboring node $\vec{r}+\vec{c}^j$ (see Fig. \ref{fig:TransitionProcess}). Thus, the dynamics can be summarized by the stochastic difference equation \begin{equation}
    s^j_{\vec{r}+\vec{c}^j}(\tau+1)={s^{\mathcal{I}}_{\vec{r}}}^j(\tau).
    \label{dyneq}
\end{equation}

The transition probability $P(\vec{s}_{\vec{r}}\rightarrow\vec{s}^{\mathcal{I}}_{\vec{r}})$ between node state $\vec{s}_{\vec{r}}$ and $\vec{s}^{\mathcal{I}}_{\vec{r}}$, defines the system dynamics which depends only on the initial and final states, as occurs in the description of the stochastic dynamics of systems that satisfy detailed balance. Thus, the transition probability is proportional to $\exp\{- H(\vec{s}_{\vec{r}},\vec{s}^{\mathcal{I}}_{\vec{r}})\}$, where  $H(\vec{s}_{\vec{r}},\vec{s}^{\mathcal{I}}_{\vec{r}})$ is reminiscent of the relative dimensionless ``energy'' between the two states. In this paper we consider two independent processes that rule the transitions from one state to another, i.e.,   $H(\vec{s}_{\vec{r}},\vec{s}^\mathcal{I}_{\vec{r}})=H_\textrm{pers}(\vec{s}_{\vec{r}},\vec{s}^\mathcal{I}_{\vec{r}})+H_\textrm{alig}(\vec{s}_{\vec{r}},\vec{s^\mathcal{I}}_{\vec{r}})$. The first term involves only the on-site occupancy states $\vec{s}_{r}$, $\vec{s}^{\mathcal{I}}_{\vec{r}}$, and models  \emph{persistent} dynamics. For this we choose time and space-discrete overdamped Langevin dynamics with exponentially correlated Gaussian noise, corresponding to the type of active motion introduced in \cite{KMDL14}, given by \cite{nava2017cellular}
\begin{subequations}
\begin{equation}
H_\textrm{pers}(\vec{s}_{\vec{r}},\vec{s}^\mathcal{I}_{\vec{r}})=-\alpha\Biggl[\sum_{\ell=1}^6\sum_{j=1}^6\left({s_{\vec{r}}^{\mathcal{I} \ell}}\vec{c}^{\ell}\cdot s_{\vec{r}}^j\vec{c}^j\right)\Biggr],
\end{equation}
where $\alpha>0$ is a parameter that characterizes the persistence of motion. For a given occupation state $\vec{s}_{\vec{r}}$ and finite $\alpha$, the probability of the transition $\vec{s}_{\vec{r}}\rightarrow\vec{s^{\mathcal{I}}}_{\vec{r}}$ is maximum for $\vec{s}^{\mathcal{I}}_{\vec{r}}=\vec{s}_{\vec{r}}$, i.e. the state $\vec{s}_{\vec{r}}$, persists. In the limit $\alpha\rightarrow0$ the transition among velocity channels are equally probable and thus not persistent motion is observed.

The second process involves the occupancy states of the neighboring cells and defines a nematic velocity alignment interaction \cite{nava2017extracting}, among the particles in node $\vec{r}$ and the particles in nearest neighboring nodes, namely 
\begin{equation}
H_\textrm{alig}(\vec{s}_{\vec{r}},\vec{s}^{\mathcal{I}}_{\vec{r}})=-\beta\Biggl[\sum_{\vec{r}^{\mathcal{I}}\in\mathcal{N}_{\vec{r}}}\sum_{\ell=1}^6\sum_{j=1}^6\left(\vec{c}^{\ell}\cdot\vec{c}^j\right)^2 s_{\vec{r}}^{\mathcal{I} \ell}s_{\vec{r}^{\mathcal{I}}}^{j}\Biggr]
\end{equation}
\end{subequations}
where $\beta>0$ controls the sensitivity towards the nematic interaction and $\mathcal{N}_{\vec{r}}$ refers to the set of nearest neighboring nodes $\vec{r^{\mathcal{I}}}$ of node $\vec{r}$.
Thus
\begin{equation}\label{trans-real}
P(\vec{s}_{\vec{r}}\rightarrow\vec{s}^{\mathcal{I}}_{\vec{r}})=\frac{1}{Z}\exp\bigl\{-H(\vec{s}_{\vec{r}},\vec{s}^\mathcal{I}_{\vec{r}})\bigr\}\delta_{n(\vec{s}_{\vec{r}}),n(\vec{s}^\mathcal{I}_{\vec{r}})}
\end{equation}
with 
$\delta_{n(\vec{s}_{\vec{r}}),n(\vec{s}^\mathcal{I}_{\vec{r}})}$ guaranteeing particle conservation during the transition process at each node, and $Z$ being the normalization constant, reminiscent of a partition function \begin{equation*}
Z=\sum_{\vec{s}^\mathcal{I}_{\vec{r}}\in\mathcal{E}}\exp\bigl\{-H(\vec{s}_{\vec{r}},\vec{s}^\mathcal{I}_{\vec{r}})\bigr\}\delta_{n(\vec{s}_{\vec{r}}),n(\vec{s}^\mathcal{I}_{\vec{r}})}.
\end{equation*} Notice that, since the sum is over all 6-tuples with values either 0 or 1 (i.e. all elements of $\mathcal{E}$), velocity channels can never be occupied by more than one particle.  In the case of vanishing $\alpha,\beta$, the transition probability between any pair $\vec{s}_{\vec{r}},\vec{s}^{\mathcal{I}}_{\vec{r}}$ is uniform, thus neither persistence nor alignment occurs, and the dynamics is similar to the stochastic dynamics of a simple random walk. 

\section{\label{sect:numerics}Numerical analysis of the model}
Computer simulations of the LGCA model were performed on a hexagonal lattice of $L\times L$ nodes with $L=120$, considering periodic boundary conditions. The initial state for all simulations was chosen to be a spatially homogeneous state on the average, i.e., with equal average density $\rho=N/L^2$ at every node, $N$ being the total number of particles in the system, and with every velocity channel having equal probability to be occupied. In the present study we kept $\rho=0.2$ for the numerical simulations, which corresponds to a dilute regime. We focus our analysis in this rather nontrivial regime, since collective behavior in the crowded regime is well understood.

\begin{figure}
\includegraphics[width=0.5\columnwidth]{lowahighb.pdf}
\includegraphics[width=0.5\columnwidth]{highahighb.pdf}\\
\hspace{-0.9cm}\includegraphics[width=0.5\columnwidth,trim=0 0 0 0, clip=true]{highalowb.pdf}
\qquad\qquad\includegraphics[width=0.25\columnwidth]{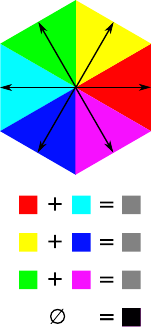}\qquad

\caption{Top and lower left.- Simulation snapshots obtained from LGCA model. Simulations consider interactions with the six neighbors of each node according to the updating rules (see text) after 2000 time steps starting from disordered, homogeneous initial conditions. (top: $\alpha=0$, $\beta=7$; middle: $\alpha=5$, $\beta=7$; lower left: $\alpha =9$, $\beta=2.5$). All simulations were performed on a $120\times120$ hexagonal lattice. Lattices were initialized with a mean occupation of $\rho=0.2$. Nodes were colored according to occupied velocity channels. The color of nodes occupied by more than one particle is computed by adding the color of the occupying channels according to the color code provided.}
\label{snaps1}
\end{figure}

From computer simulations we observe the emergence of different collective patterns of motion characterized by different orientational symmetries: a single nematic band when $\alpha\rightarrow 0$ and $\beta\gg 0$ (Fig.~\ref{snaps1} top panel); a network of nematic bands when both $\alpha\approx\beta\gg0$ (Fig.~\ref{snaps1} middle panel); and polar clusters when $\alpha\gg\beta$ (Fig.~\ref{snaps1} bottom-left panel).

After letting the system reach the stationary state (approximately 1000 time steps), we measured the degree of global order of the system, characterized by two quantities: One considering the global orientational order of the system corresponding to the average direction of motion given by 
\begin{subequations}
\begin{equation}
S_F=\frac{1}{N}\left|\sum_{\vec{r}\in\mathcal{L}}\sum_{j=1}^6 e^{is_{\vec{r}}^j\theta^j}\right|;
\end{equation}
with the other measuring the appearance of nematic-like order 
\begin{equation}
S_L=\frac{1}{N}\left|\sum_{\vec{r}\in\mathcal{L}}\sum_{j=1}^6 e^{is_{\vec{r}}^j 2\theta^j}\right|
\end{equation}
\end{subequations}
indicating directional order independently of the orientation of motion; where $|\cdot|$ denotes the modulus of a complex number, $\theta^j$ is the angular component of the $j$-th velocity channel, $i$ is the imaginary unit, and $N$ is the total number of particles in a specific realization of the system, given by $\sum_{\vec{r}\in\mathcal{L}}\sum_{j=1}^6 s_{\vec{r}}^j$.
\begin{figure}[th]
\includegraphics[width=\columnwidth,trim=20 3 45 18, clip=true]{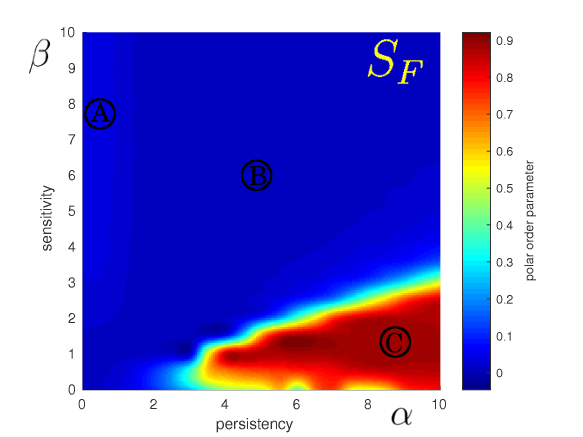}
\includegraphics[width=\columnwidth,trim=20 3 45 18,clip=true]{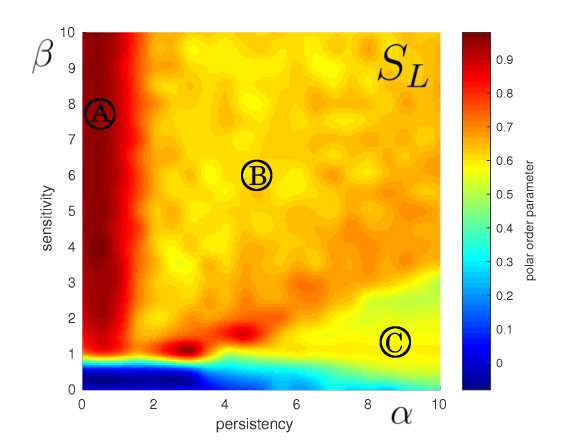}
\caption{Simulation results of persistent, nematic-aligning particles. 
Top.- Polar order parameter values as a function of persistency $\alpha$ and sensitivity $\beta$, with density $\rho=1/5$.
Bottom.- nematic order parameter values as a function of persistency $\alpha$ and sensitivity $\beta$, with density $\rho=1/5$. These plots show three main areas (A,B, and C), with different global directional and orientational order parameter values, and different macroscopic patterning. A small area ($\alpha\sim 0$, $\beta\sim 0$) with no order is the random walk limit.}\label{nemorder-params}
\end{figure}
Although simulations were performed with the specific particle density $\rho=0.2$, different densities lead to qualitatively the same stationary behavior. The qualitative density independence becomes obvious from a linear stability analysis, which shows that the behavior close to the homogeneous state only depends on the product of density and the nematic sensitivity $\beta$/persistence $\alpha$, not on their individual values; see Section IV below, and \cite{bussemaker1997mean}. This behavior is not exclusive to discrete models, see also \cite{peruani2008mean}. 

Simulation results are reported in Fig. \ref{nemorder-params} where the three distinct collective behaviors are mapped in the phase diagram $\beta $-$\alpha$:
A region where the system exhibits high nematic order $S_{L}\lesssim1$ but low polar order $S_F\sim0$ for small persistence and high sensitivity, denoted in Fig. \ref{nemorder-params} (top and bottom) with \textbf{A}, corresponding to the system snapshot at the top-left panel of Fig. \ref{snaps1}. A large region for intermediate values of $\alpha$ and $\beta$, where polar order remains low, but the nematic one diminishes, is marked with \textbf{B} (top-right snapshot in Fig.~\ref{snaps1}). A region where polar alignment rises and nematic order lowers, for large values of $\alpha$, this region is marked with \textbf{C} (bottom-left snapshot in Fig.~\ref{snaps1}). Absence of polar and nematic order is observed for small values of $\alpha$ and $\beta$.

\begin{figure}
\includegraphics[width=\columnwidth]{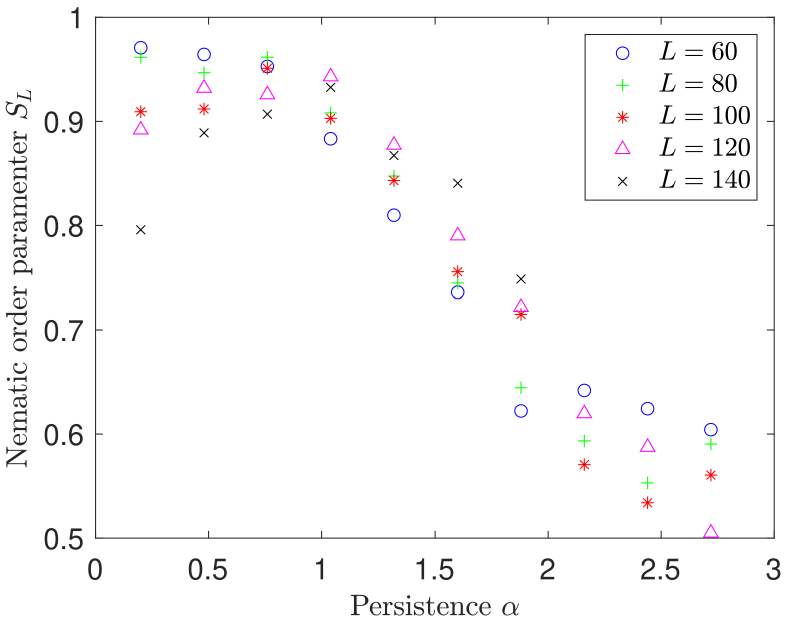}
\caption{Phase transition from region A to region B with varying system sizes. The phase transition between the globally nematically aligned state (A, Fig.~\ref{nemorder-params}) to the frustrated state (B, Fig.~\ref{nemorder-params}) produced by incrased persistence $\alpha$ was observed through computer simulations while varying the value of the system size $L$. Particle density was fixed at $\rho=0.2$, and sensitivity at $\beta=4$. The system was simulated for $1000$ time steps, and the value of the nematic order parameter was recorded at the end of the simulation. The values reported are averages over $100$ simulations.}\label{sys-size}
\end{figure}

To study the effect of system size on the novel transition from global nematic order with low persistence to a frustrated, local nematic order with high persistence, we simulated the model near the phase transition (corresponding to the boundary between regions A and B in Fig.~\ref{nemorder-params}), for several system sizes $L$. As can be observed in Fig.~\ref{sys-size}, with increasing system size the gap between the globally ordered state and the partially ordered state seems to increase. This may be due to the fact that, for greater system sizes, there is more space available for different bands to form and to recruit particles, while for small system sizes, only a few bands can form due to the overall width of the bands.

\section{\label{sect:analysis}Mean field analysis}
We also performed a linear stability analysis of the homogeneous phase against the breaking of orientational symmetry. 
To simplify the analysis, we assumed a diluted system such that the density is low enough that the transition among occupation states can be approximated by an effective single-particle transition probability, as well as a mean-field approximation (Sto{\ss}zahlansatz).
The mean-field theory of our analysis is built by considering the expectation values of Eq.~\ref{dyneq}, which under the assumption of a dilute system, results  in the partial difference equation
\begin{equation}
f_{\ell}(\vec{r}+\vec{c}^\ell,\tau+1)=\biggl(\sum_{n=1}^6 f_{n}(\vec{r},\tau)\biggr)T_{\ell},
\label{microdyn}
\end{equation}
 where $f_{\ell}(\vec{r},\tau)\in[0,1]$ is the expected value of the occupation of channel $\ell$ at node $\vec{r}$ and time $\tau$, i.e., $f_\ell\coloneqq\left\langle s^\ell_{\vec{r}}\right\rangle$, and $T_{\ell}$ is the probability that a particle occupies the channel $\ell$, given by
 \begin{multline}
 T_{\ell}(\vec{r},\tau)=\frac{1}{Z}\exp\Biggl[\alpha\sum_{j=1}^6\left(\vec{c}^{\ell}\cdot\vec{c}^j\right)f_j(\vec{r},\tau)\\ +\beta\sum_{p=1}^6\sum_{j=1}^6\left(\vec{c}^{\ell}\cdot\vec{c}^j\right)^2f_j(\vec{r}+\vec{c}^p,\tau)\Biggr],
 \label{sing-prob}
 \end{multline}
 which is obtained from Eq.~\eqref{trans-real} after the mean field approximation, by considering the reorientation of a single particle, rather than the complete configuration, and assuming no interference among the reorientation of particles within a single node. Thus Eq. \eqref{microdyn} together with \eqref{sing-prob}, defines the self-consistency equation of the mean-field theory.
 
In the homogeneous phase, the mean particle density is translationally invariant on $\mathcal{L}$ and thus $\rho(\vec{r})=\rho$, from this symmetry we have that average number of particles at each cell $\vec{r}$, $\sum_{\ell=1}^6 f_\ell$, is equal to $6\rho$.
 The mean-field homogeneous stationary solutions, if any, must satisfy $f_{\ell}(\vec{r}+\vec{c}^j,\tau+1)=f_{\ell}(\vec{r},\tau)$ for all $\vec{r}$ and $\tau$, thus  Eq.~\ref{microdyn} is transformed into
\begin{equation}
f_{\ell}=\left(\sum_{n}f_n\right)T_{\ell}.
\label{sseq}
\end{equation}

\begin{figure} \includegraphics[width=\columnwidth]{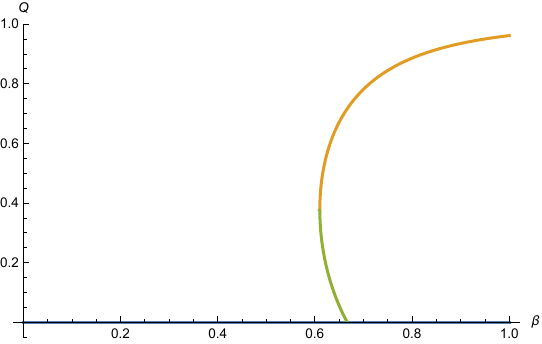}
\caption{Numerical bifurcation diagram showing the steady state nematic order director $Q$ with changing sensitivity $\beta$. The state $Q_+$ is shown in yellow, the state $Q_-$ is show in green.}
 \label{steadys}
\end{figure}

We proceed further as follows: without loss of generality, we assume the case of nematic alignment among particles, and choose the nematic axis as the one that contains the channels directions $\vec{c}^1$ and $\vec{c}^4$ thus 
\begin{subequations}
 \begin{align}
  f_1 & =f_4, \\
  f_2 & =f_3=f_5=f_6.
 \end{align}
 \label{nemsscond}
\end{subequations}
Under this assumption, the alignment term does not contribute to $T_\ell$ and thus the single particle probability reduces to
\begin{equation}
 T_{\ell}=\frac{1}{Z}\exp\left[6\beta \left(f_1-f_2\right)\cos\left(\frac{2\pi}{3}(\ell-1)\right)\right].
 \label{nemssprobs}
\end{equation}
Except for the trivial (disordered) steady state, Eq.~\eqref{sseq} together with the assumptions Eqs.~\ref{nemsscond} and \ref{nemssprobs} cannot be solved exactly, and thus we resort to numerical calculations to find the steady states with non-vanishing nematic order parameter 
\begin{multline}
 Q\coloneqq\Bigglb[\Biggl\{\sum_{j=1}^6 f_j\cos\left[\frac{2\pi}{3}(j-1)\right]\Biggr\}^2+\\
 \Biggl\{\sum_{j=1}^6f_j\sin\left[\frac{2\pi }{3}(j-1)\Biggr]\right\}^2\Biggrb]^{1/2}.
\end{multline}
For a fixed density $\rho$, the disordered state ($Q=0$) is stable only for low values of the sensitivity $\beta$. As this is increased, nematic-ordered states emerge. For these ordered states, $f_1>f_2$ (Fig.~\ref{steadys}). There are two nematically ordered states, one with $f_1\gg f_2$, which we will name $Q_+$, corresponding to a high value of the nematic director (nematic alignment along a single axis), and another with $f_1>f_2$, but with $4f_2>2f_1$, named $Q_-$, corresponding to a low value of the nematic director (a higher proportion of mass along two different axes).

We now investigate the stability of each of the steady states. Following \cite{bussemaker1997mean} and \cite{Deutsch2005}, we linearize Eq.~\ref{microdyn} around one of the steady states and, performing a discrete Fourier transform, yields the equation
\begin{equation}
\hat{f}_{\ell}\left(\vec{k},\tau+1\right)=\sum_{j=1}^b\Gamma_{\ell,j}\left(\vec{k}\right)\hat{f}_j\left(\vec{k},\tau\right),
\end{equation}
where $\hat{\cdot}$ represents the discrete Fourier transform, and $\Gamma_{\ell,j}\left(\vec{k}\right)$ is called the Boltzmann propagator, defined as 
\begin{equation}
\begin{split}
\Gamma_{\ell,j}(\vec{k})&=e^{-\frac{2\pi i}{L}\vec{k}\cdot\vec{c}^{\ell}}\left[\delta_{\ell,j}+\left.\frac{\partial C_{\ell}}{\partial f_j(\vec{r},\tau)}\right|_{\mathrm{st.s}}\right.\\&\left.+\sum_{p=1}^6e^{\frac{2\pi i}{L}\vec{k}\cdot\vec{c}^p}\left.\frac{\partial C_{\ell}}{\partial f_j(\vec{r}+\vec{c}^p,\tau)}\right|_{\mathrm{st.s}}\right],
\end{split}
\label{boltz-prop}
\end{equation}
where $\vec{k}$ is the wave vector, $L$ is the number of lattice sites in each velocity channel direction, and the change of occupation numbers is defined as $C_{\ell}=f_{\ell}(\vec{r}+\vec{c}^j,\tau+1)-f_{\ell}(\vec{r},\tau)$. The eigenvectors and eigenvalues of the Boltzmann propagator gives information about the stability of the steady states.

We start with the stability of the disordered state, $f_1=f_2=f_3=f_4=f_5=f_6=\rho$. 
Using Eqs. \ref{microdyn}, \ref{nemssprobs}, and \ref{boltz-prop}, we find that at this steady state, the Boltzmann propagator is given by
\begin{equation}
\begin{split}
\Gamma_{\ell,j}(\vec{k})&=e^{-\frac{2\pi i}{L}\vec{k}\cdot\vec{c}^{\ell}}\left\{\frac{1}{6}+\alpha\rho\left(\vec{c}^{\ell}\cdot\vec{c}^j\right)\right.\\&\left.+\beta\rho\left[2\left(\vec{c}^{\ell}\cdot\vec{c}^j\right)^2-1\right]\sum_{p=1}^3\cos\left(\frac{2\pi}{L}\vec{k}\cdot\vec{c}^p\right)\right\}.
\end{split}
\end{equation}

\begin{figure}
\centering
\includegraphics[width=\columnwidth]{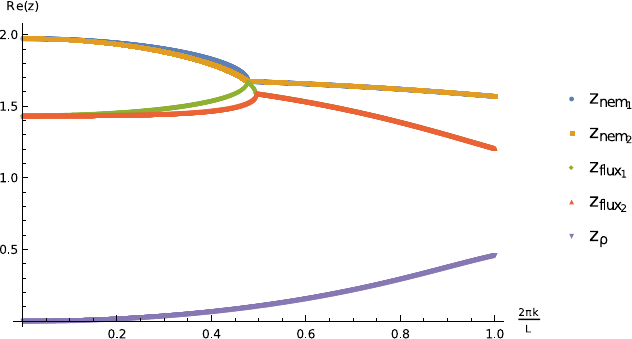}
\includegraphics[width=\columnwidth]{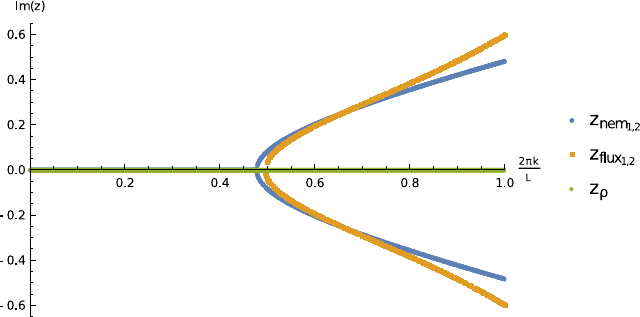}
\caption{Eigenvalue logarithms $z$ corresponding to the disordered state, as a function of the wave number $\vec{k}$, real (top) and imaginary (bottom) parts. The eigenvalue logarithm was evaluated along the vector parallel to the $x$ axis, $\vec{k}=(k,0)$.  
$z(k)$ corresponding to $\lambda_{\mathrm{tot.flux}}$ is not shown as $\lambda_{\mathrm{tot.flux}}=0$ for every $k$. Parameter values were set at $\rho=0.2$, $\alpha=7$, $\beta=4$.}
\label{zcases}
\end{figure}

We start by looking at the homogeneous case $\vec{k}=(0,0)$. 
First, we find an eigenvalue $\lambda_{\rho}=1$ with the corresponding eigenvector $(1,1,1,1,1,1)$, which is related to mass conservation. Then we find an eigenvalue $\lambda_{\mathrm{tot.flux}}=0$ with eigenvector $(-1,1,-1,1,-1,1)$, which represents the flux along all lattice axes. We find a twofold degenerate eigenvalue $\lambda_{\mathrm{flux}_{1,2}}=3\alpha\rho$ with eigenvectors $(1,0,-1,-1,0,1)$ and $(-1,-1,0,1,1,0)$, representing fluxes along two lattice axes. This explains the transition to the polarly-aligned state (Fig.~\ref{nemorder-params}, region \textbf{C}) Finally, we encounter a fourth, twofold degenerate eigenvalue $\lambda_{\mathrm{nem}_{1,2}}=9\beta\rho$, with eigenvectors $(-1,0,1,-1,0,1)$ and $(-1,1,0,-1,1,0)$, which represents nematic alignment along two lattice axes. This instability relates to the transition to the completely nematically-aligned state (Fig.~\ref{nemorder-params}, region \textbf{A}). It can be seen that the homogeneous pattern becomes unstable when $3\alpha\rho=1$, in which case the flux components of the perturbation grow, and when $9\beta\rho=1$, in which case the nematic components of the perturbation grow. 

For non-homogeneous perturbations ($\vec{k}\neq 0$), following \cite{bussemaker1997mean}, it is convenient to define an eigenvalue logarithm, defined as $z\left(\vec{k}\right)=\ln\left[\lambda\left(\vec{k}\right)\right]$, so that perturbations grow like $e^{z\left(\vec{k}\right)}$. Modes are unstable when $\mathrm{Re}\left[z\left(\vec{k}\right)\right]>0$ and stable when $\mathrm{Re}\left[z\left(\vec{k}\right)\right]<0$. Moreover, perturbations travel with nonzero speed when the imaginary part $\mathrm{Im}\left[z\left(\vec{k}\right)\right]\neq 0$. 

When $\beta\gg \alpha$ both nematic modes are unstable, while all other modes are stable for every value of the wave vector $\vec{k}$. All $z$ are real, which indicates no net transport takes place (purely diffusive behavior). This is agrees qualitatively to continuous migration models with nematic alignment \cite{peruani2008mean}.

On the other hand, when $\alpha \gg \beta$ the case is completely analogous to the case of polar alignment studied in \cite{bussemaker1997mean}, where density and flux modes travel with a certain velocity and a restricted finite wavelength.

Finally, when both $\alpha$ and $\beta$ have high values (Fig.~\ref{zcases}), all modes are unstable. As in the first case, nematic modes dominate. After a critical wave number, both nematic modes and both flux modes merge, while at the critical point, nematic modes and a flux mode coexist. The density mode remains separate from all other modes. Meanwhile, both nematic and both flux modes have non-zero imaginary parts starting from the critical point. This indicates that nematic and flux modes travel with finite speed. The density mode remains real for every wave number, indicating that traveling nematic and polar structures form, but no net mass transport occurs. 

\begin{figure}
\centering
\includegraphics[width=\columnwidth]{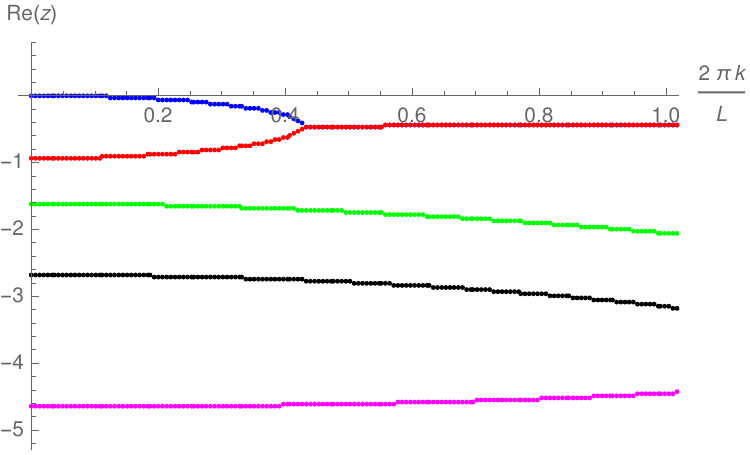}

\includegraphics[width=\columnwidth]{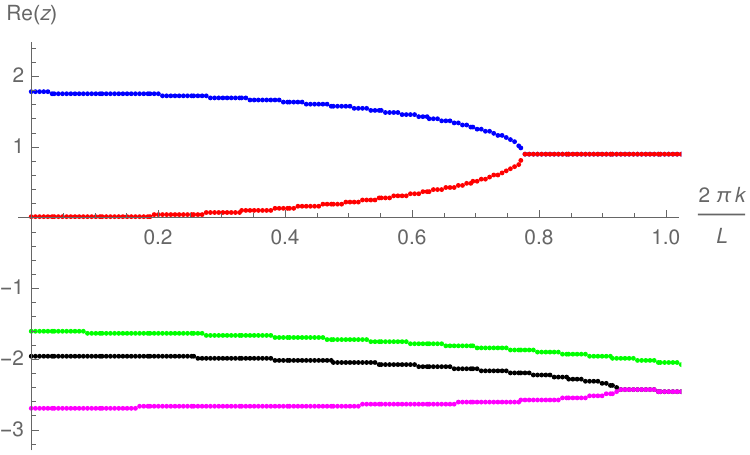}

\caption{Eigenvalue logarithms $z$ corresponding to the ordered state $Q_+$, as a function of the wave number $\vec{k}$. The persistence parameter was set to $\alpha=0.3$ (top) and $\alpha=5$ (bottom). Other parameters were set at $\rho=0.2$ and $\beta=0.8$ in both cases. The eigenvalue logarithm was evaluated along the vector parallel to the $x$ axis, $\vec{k}=(k,0)$.}
\label{zcasesnem}
\end{figure}

Conversely, the Boltzmann propagator at the homogeneous, nematically ordered steady states is
\begin{equation}
\begin{split}
 \Gamma_{\ell,j}\left(\vec{k}\right)&=e^{-\frac{2\pi i}{L}\vec{k}\cdot\vec{c}^{\ell}}T_{\ell}^{\mathrm{nem}}\left\{1+\left(2f_1+4f_2\right)\left[\alpha\left(\vec{c}^{\ell}\cdot\vec{c}^j\right)\right.\right.\\&
 +\beta\left(2\left(\vec{c}^{\ell}\cdot\vec{c}^j\right)^2-1-\cos\left(\frac{2\pi j}{3}\right)\frac{E_1-E_2}{E_1+2E_2}\right)\\&\times\left.\left.\sum_{p=1}^3\cos\left(\frac{2\pi}{L}\vec{k}\cdot\vec{c}^p\right)\right]\right\},
 \end{split}
\end{equation}
where we define the non-normalized weights $E_{\ell}\coloneqq\exp\left[6\beta \left(f_1-f_2\right)\cos\left(\frac{2\pi \ell}{3}\right)\right]$.
Numerically, we find that, for $\alpha=0$, the nematic steady state $Q_+$ is stable, while $Q_-$ is unstable. 
This is not surprising, since in the first case there is nematic alignment along one axis, while in the second there is nematic alignment along two axes. Also in agreement with the findings in \cite{baskaran2008hydrodynamics} we find that, while for low non-zero persistence $Q_+$ remains stable (Fig.~\ref{zcasesnem}, top), large non-zero persistence $\alpha$ destabilizes the nematic steady state $Q_+$, creating unstable polar and nematic modes (Fig.~\ref{zcasesnem}, bottom). This analysis explains the observed transition from region \textbf{A} with complete nematic alignment, to region \textbf{B} with partial nematic alignment (see Fig.~\ref{nemorder-params}) by increasing the persistence $\alpha$.

\section{\label{sect:conclusions}Summary and discussion}
In this work we have introduced a \emph{generic} model of active matter that takes into account aligning interactions among self-propelled Brownian particles We show that the collective dynamics that emerge from local-nematic-aligning interactions, is strongly influenced by the persistence of the motion of the individuals. Our model is based on a lattice-gas cellular automata approach which is amenable to numerical simulations and analytical treatment in the mean-field approximation, in contrast to other agent-based models. An advantage of the model is that the individual motility and the aligning collective interactions contribute to the dynamics in a well-separated manner, allowing the analysis of the competition between them, as may happen in realistic systems. For instance, in the Vicsek model both elements are intrinsically interconnected through the noise amplitude. In other models, persistence and interaction strength are coupled through the Péclet number \cite{zottl2023modeling}. Other agent-based models can be used to consider individual and collective effects separately, however at the cost of mathematical simplicity \cite{wortel2021local}.

Two parameters define the phases of collective motion, the sensitivity to the alignment interaction, $\beta$, and the intrinsic persistence of motion of the individuals, $\alpha$. Our analysis at fixed particle density, indicates the existence of four distinct phases with different order parameters, corresponding to different modes of collective motion. Interestingly, we found phases where collective motion is frustrated by the persistence of motion of the individuals.
In the case of low sensitivity to alignment and low persistence ($\beta\ll1$ and $\alpha\ll1$ respectively), a stable homogeneous phase with no polar nor nematic order appears. 
At a high sensitivity $\beta$, but low persistence $\alpha$, the disordered phase becomes unstable and a globally ordered nematic state emerges. Conversely, for low sensitivity $\beta$ and high persistence $\alpha$, a state with global polar order is observed. Finally, when both sensitivity $\beta$ and persistence $\alpha$ are high, a state with partial nematic order emerges.

The influence of persistence is conspicuous in the collective dynamics, as we have  mathematically found steady states that exhibit nematic order at sufficiently high values of the sensitivity: one with high nematic order, a second with low nematic order, and a third one with no nematic order. We would like to point out that only two out of these three states can be observed computationally, since the disordered state and the state with high nematic order are stable for certain parameter regimes. On the other hand, the steady state with low nematic order is never stable, and therefore not observable.

Our model provides evidence that persistence can act as a control parameter regarding macroscopic patterns formed by self-propelled particles. This is particularly interesting as it suggests that particle preference to move along a certain direction can be tuned to achieve certain structural outcomes, which could be used for the design of active materials. Furthermore, our results indicate that in systems where particles exhibit nematic alignment, persistence can be a decisive factor in determining which type of order dominates. The spatial patterns we observe in simulations are similar to those observed experimentally in \cite{nishiguchi2017long}, which report local, but not global nematic order, for a population of persistent \emph{E. coli} bacteria. Even though they do observe global nematic order for very high cell densities, this could be a result of hydrodynamic interactions with the surrounding fluid or steric repulsion among cells, which we did not consider here. This behavior is analogous to that in microtubules moved by kinesin motors with tunable activity through varying ATP concentration \cite{memarian2021active}, and the nematic networks formed by \emph{C. elegans} \cite{sugi2019c}. Our results also agree with those found in \cite{thutupalli2015directional}, where it was observed that populations of \emph{M. xanthus} bacteria formed 2D sheets (analogous to Fig.~\ref{snaps1} top) as well as streams (analogous to Fig.~\ref{snaps1} middle). Importantly, our model not only agrees with patterns formed by bacterial systems. Our minimal model also reproduces patterns observed in human tumor cells, where globally nematic ordering is commonly observed along with regions of high local nematic order, but bounding regions with different ordering, as well as polar flocking structures \cite{barberis2024self}.Thus, our results could shed light on the mechanisms underlying the formation and cohesion of biological swarms.

Our research thus provides insight into the fundamental mechanisms driving pattern formation in active matter systems, highlighting the importance of persistence in nematically aligning particles, which is often disregarded, but which could have significant impacts on the possibility of achieving global ordering within the population \cite{shi2014instabilities,zottl2016emergent,snezhko2016complex,caprini2023flocking}.

Our model still presents open questions which will be addressed in upcoming work. To begin with, the critical exponents at the five different phase transition among the four observed steady states can be obtained explicitly. 
These exponents define a rescaling of the order parameters and the system size which, in turn, allow to characterize the order of the phase transition among these states. These exponents would allow to transform our results (Fig.~\ref{sys-size}) and to better understand the critical nature of the phase transition \cite{baglietto2012criticality}. Furthermore, here we have only characterized the patterns formed in a square lattice with periodic boundary conditions. It remains to be determined whether such patterns are affected by changes in the aspect ratio of the simulation area and the boundary conditions, in a similar fashion to continuous agent-based models.

The model introduced in this work is susceptible to straightforward generalizations in order to study physically relevant situations. In this paper we have considered persistent particles with exponentially decaying velocity autocorrelations \cite{nava2017cellular}. It would be interesting to explore whether similar trends are observed when either autocorrelations become more slowly decaying (for example, as a power law), or jump lengths are fat-tailed (L\'evy) distributed. Additionally, it is an open problem whether intrinsic movement preference affects the population-level behavior of a system of interacting particles independently of the nature of the inter-particle interaction, or if such effects are only allowed by certain interaction types. Furthermore, particles in our model do not show steric interactions, in the sense that nearby particles may still move at oblique angles, albeit with low probability. This can be prevented by including zero-velocity channels, whose occupation probability grows with particle density and orientation mismatch, in a similar fashion to \cite{ilina2020cell}. Changes in interactions defining reorientations can be straightforwardly implemented by changing the Hamiltonian $H(\vec{s}_{\vec{r}},\vec{s}^{\mathcal{I}}_{\vec{r}})$, as described in \cite{deutsch2021bio}. Environmental effects can also be considered by defining a (not necessarily time-independent) scalar or vector field on $\mathcal{L}$ which impacts transition probabilities (such as the extracellular matrix field in \cite{ilina2020cell}). Additionally, alien or disruptive particles (corresponding to defects in inert crystals) could be introduced by following \cite{mente2015analysis,reher2017cell}, whereby particles are no longer identical, such that individual identities and behaviors can be assigned to any or all of them.

Finally, here we have explored some consequences of the effects of the
individual behavior (persistence in this case) on the macroscopic
collective properties of the system. The converse problem, which would
look at the effect of interactions on the motility of individual
particles, has only been studied very recently
  \cite{ABR17,FeKo17}. It could shed light on the mechanistic origin
  of anomalous diffusion in collective motion, observed in biological systems
  \cite{harris2012generalized,ARBPHB15,nousi2021single}.

\section*{Acknowledgements}
The authors would like to thank the Centre for Information Services and High Performance Computing (ZIH) at TU Dresden, as well as the Mathematics Department at UNAM for providing an excellent infrastructure.

\end{document}